\documentclass[12pt]{iopart}
\usepackage{hyperref}

\usepackage{siunitx}
\usepackage{fontawesome}
\usepackage{calligra}
\usepackage{graphicx}
\usepackage{caption}
\DeclareMathAlphabet{\mathcalligra}{T1}{calligra}{m}{n}
\DeclareFontShape{T1}{calligra}{m}{n}{<->s*[2.2]callig15}{}

\usepackage{scalerel}
\usepackage{tikz}
\usetikzlibrary{svg.path}

\definecolor{orcidlogocol}{HTML}{A6CE39}
\tikzset{
  orcidlogo/.pic={
    \fill[orcidlogocol] svg{M256,128c0,70.7-57.3,128-128,128C57.3,256,0,198.7,0,128C0,57.3,57.3,0,128,0C198.7,0,256,57.3,256,128z};
    \fill[white] svg{M86.3,186.2H70.9V79.1h15.4v48.4V186.2z}
                 svg{M108.9,79.1h41.6c39.6,0,57,28.3,57,53.6c0,27.5-21.5,53.6-56.8,53.6h-41.8V79.1z M124.3,172.4h24.5c34.9,0,42.9-26.5,42.9-39.7c0-21.5-13.7-39.7-43.7-39.7h-23.7V172.4z}
                 svg{M88.7,56.8c0,5.5-4.5,10.1-10.1,10.1c-5.6,0-10.1-4.6-10.1-10.1c0-5.6,4.5-10.1,10.1-10.1C84.2,46.7,88.7,51.3,88.7,56.8z};
  }
}

\newcommand\orcidicon[1]{\href{https://orcid.org/#1}{\mbox{\scalerel*{
\begin{tikzpicture}[yscale=-1,transform shape]
\pic{orcidlogo};
\end{tikzpicture}
}{|}}}}

\DeclareRobustCommand{\uvec}[1]{{%
     \csname uvec#1\endcsname
   \else
    \bm{\hat{\mathbf{#1}}}%
   \fi
}}

\begin{document}

\title[Modeling damped oscillations  in U-tubes  via Easy JavaScript Simulations]{A comprehensive modeling and experimental approach for damped oscillations  in U-tubes  via Easy JavaScript Simulations}

	\author{Fredy A. Orjuela\orcidicon{0000-0003-1556-504X}}
        \address{Universidad de los Andes, Bogot\'a, Colombia}
        \ead{fa.orjuela@uniandes.edu.co}

        \author{J. E. Garc\'ia-Farieta\orcidicon{0000-0001-6667-5471}}
        \address{Instituto de Astrof\'isica de Canarias, s/n, E-38205, La Laguna, Tenerife, Spain.\\Departamento de Astrof\'isica, Universidad de La Laguna, E-38206, La Laguna, Tenerife, Spain}
        \ead{jorge.farieta@iac.es}
        
        \author{H\'ector J. Hort\'ua\orcidicon{0000-0002-3396-2404}}
        \address{Grupo Signos, Departamento de Matem\'aticas, Universidad El Bosque, Bogot\'a, Colombia.}
        \address{Maestr\'ia en Ciencia de Datos, Universidad Escuela Colombiana de Ingenier\'ia Julio Garavito,  Bogot\'a, Colombia.} 
        \ead{hhortuao@unbosque.edu.co}	
        
        \author{E. Munevar\orcidicon{0000-0002-0578-7717}}
        \address{Grupo Fisinfor, Universidad Distrital Francisco Jos\'e de Caldas, Bogot\'a, Colombia.}
        \ead{emunevare@udistrital.edu.co}

\begin{abstract}
In recent years,  science simulations have become popular among educators due to their educational usefulness, availability, and potential for increasing the students' knowledge on scientific topics. In this paper, we introduce the implementation of a user-friendly simulation based on Easy Java/JavaScript Simulations (EJS) to study the problem of damped oscillations in U-tubes\textbf{\includegraphics[height=10pt,width=10pt]{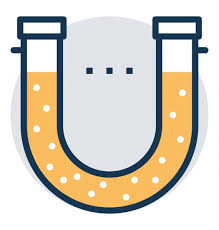}}\href{https://teaching-physics.github.io/liquid-oscillations-Utube/}{EJS}. Furthermore, we illustrate various advantages associated with the capabilities of EJS in terms of design and usability in order to encourage teachers to use it as an educational supplement to physics laboratories\href{https://github.com/teaching-physics/liquid-oscillations-Utube}{\faGithub}.
\end{abstract}

\section{Introduction}
Nowadays, virtual laboratories have become an important element for education~\cite{Monther}. These environments incorporate  visualization and interaction capabilities   under well-designed software that offers an innovative pedagogical approach to guide teaching and transform learning. When institutes lack sufficient technical equipment or resources, adequate facilities, or teaching training, among others~\cite{Rahman}, virtual laboratories offer a number of advantages over physical classrooms. In addition, these tools permit students to prepare and conduct laboratory experiments at their leisure, control multiple experiment parameters, modify conditions that cannot always be controlled in physical facilities, and monitor the resulting readings and outputs. Moreover, these scenarios allow students to experience a real-life environment and learn more visually, which is a substantial and effective learning method~\cite{Elmqaddem}. Since most of  virtual laboratories are web-based, students can access the platforms from any device, from any location, and at any time. As a matter of fact, during the COVID-19 pandemic, which imposed a new reality on many sectors, including education, virtual environments turned into a valuable resource for numerous institutes. Today, hybrid classrooms are considered a great alternative to improve the traditional method of learning by combining the best of both the real and virtual lab scenarios~\cite{Monther,Hut}. For the former, thanks to the personal student-teacher interaction, the student will be immersed in an atmosphere where aptitudes such as knowledge acquisition, ability training, and emotional edification are developed during the learning process~\cite{Cukurova,Laru}. In addition, students can acquire  experience related to management and problem-solving experimental techniques by interacting with  sensors, equipment,  weird noises and smells, and random or systematic errors. In the latter case, virtual laboratories increase the level of creativity in a virtual platform by enabling users to control environmental parameters that are difficult or impossible to set up in real-world problems,  thus filling the gap left in real classrooms. \\
Virtual laboratories have been widely developed in the physics context, providing detailed documentation and covering a variety of physics subtopics (OSP Collection~\cite{OSP}, Fisica con ordenador~\cite{fisica},   PhET~\cite{phet}, Interactive Simulations and  Easy Java Simulation~\cite{ejs1}, among others~\cite{Jorge}). One of them, Easy Java Simulation (EJS)~\cite{ejs1, ejs2}, is a free software modeling tool used in various fields due to its simplified interface for creating any simulation and the ease with which it can be executed in any web browser with standard HTML support on any computer or mobile device. Given the aforementioned discussion, we present in this paper a pedagogical EJS-based online simulation designed to provide support to instructors and students in the study of damped oscillation phenomena. This study has pedagogic applicability because it has exact analogies with, and can be considered representative of, many other phenomena of physics interest. This simulation is based on the behavior of a liquid after being displaced from its equilibrium position. It can be thought of as a starting point for the development of different physics simulations to assist and improve student learning. \\
The rest of the paper is organized as follows: section 2 briefly explains the theoretical framework behind the oscillations of a liquid column in U-tubes. Section 3 describes the experimental design and data collection for the laboratory. Furthermore, challenges and assumptions are also disclosed when these types of experiments are developed in the classroom. Section 4 shows the virtual laboratory created with Easy Java (script) Simulations, as well as a simple dashboard that visualizes the collected information of the physical phenomena. Finally, Section 5 draws some conclusions and describes future work.

\section{Theoretical description}
In real-world scenarios, most of the dissipative forces  decrease the  amplitude of an oscillating system, resulting in  damping oscillations\footnote{This is not the case, however, for collision damping in plasma vibration, where the damping takes the form of a steady draining of electrons from the vibration. In this situation, there is no decay of the amplitude associated with the plasma's collective motion.}. Let us first outline a theoretical framework  for the damped oscillations of a liquid column in U-tubes. The Newtonian formulation is essential for determining the equation of motion and describing the position, velocity, and acceleration of the liquid column at any instant of time t.\\ 

\begin{figure}[htb]
    \centering
    \includegraphics[width=1.\linewidth]{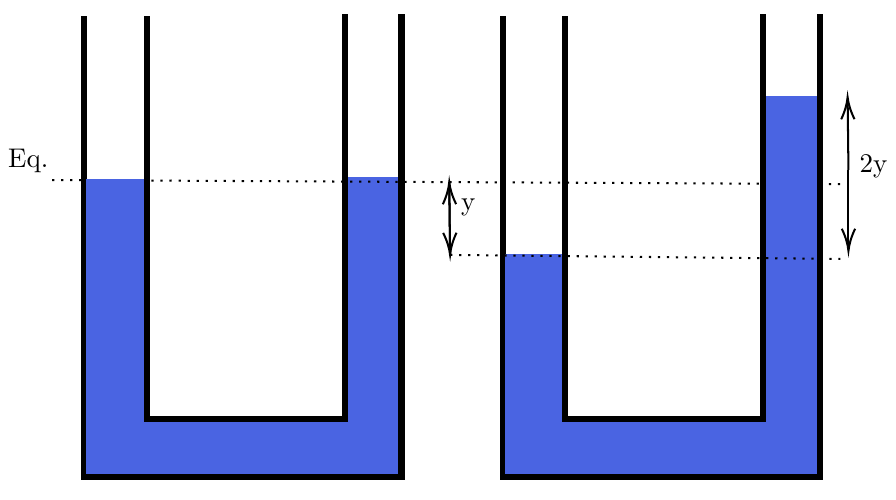}
    \caption{Visualization of the U-Tube oscillation. Left: static equilibrium. Right: Displaced column.}
    \label{fig:gui}
\end{figure}

A first approximation of the motion described by the fluid in a U-tube is given by considering only gravity and the so-called viscous frictional resistance between the fluid and the walls of the tube and neglecting the surface tension force at the upper surface of the liquid~\cite{the2}. In this case, the total force on a small element of liquid $\Delta m$ is caused by the weight (see Figure~\ref{fig:gui}) and the damping term, parameterized by $b$, that can be described as proportional to the velocity\footnote{If we consider a cylinder tube, the parameter $b$ can be written as $\pi D L \eta$, where $D$ is the diameter of the tube, and $\eta[Pa\cdot s]$ is the viscosity of the liquid~\cite{the2}. }. The equation of motion can be written as follows~\cite{the2}
\begin{equation}\label{eq:Newton1}
\rho S L\frac{d^2y}{dt^2}+b\frac{dy}{dt}+2\rho S g y =0,
\end{equation}
where $\rho$ the liquid density, $S$ the cross-section area of the tube and $L$ the liquid column  length.  Here, we have assumed a laminar  and low speed flow which implies  that  frictional forces are proportional to the velocity at which the element of liquid moves. In absence of  frictional forces, the solution of  Equation~\ref{eq:Newton1} describes the simple harmonic motion $y = A\cos(\omega_0 t+\delta)$, with  $\omega_0=\sqrt{2g/L}$.
Equation~\ref{eq:Newton1} represents a linear second-order differential equation with constant coefficients that can be solved analytically using  the ansatz  $y(t)= e^{\lambda t}$, where $\lambda$ is a parameter that contains physical information. The solution of Eq.~\ref{eq:Newton1} which describes the position of a mass element of the fluid is given by~\cite{the1} 
\begin{equation}\label{fit}
y(t)=Ae^{-\gamma t}\cos(\omega_d t+\delta),
\end{equation}
where $A$ and $\delta$ are the amplitude and the phase constant of oscillation, respectively. Both of these parameters depend on the initial conditions. The angular frequency of the oscillations is given by
\begin{equation}
\omega_d=\sqrt{\omega_0^2-\gamma^2},
\end{equation}
 being $\omega_0$ the natural frequency of the oscillations shown in the simple harmonic motion, and $\gamma=\left(\frac{b}{2\rho L S}\right)$ the damping coefficient~\cite{the2}.

\begin{figure}[htb]
    \centering
    \includegraphics[width=1.\linewidth]{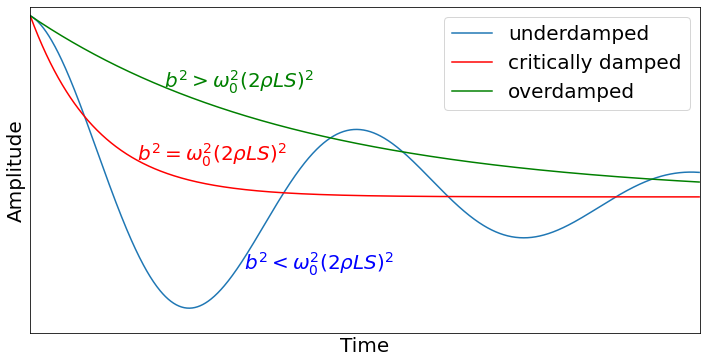}
    \caption{Amplitude versus time for three different parameter values conditions. (a) If the damping is light ($b<\sqrt{\omega_0^2(2\rho LS)^2}$), the fluid oscillates, slowly losing amplitude as the energy is dissipated by the non-conservative force. The limiting case is (b) where the damping is ($b=\sqrt{\omega_0^2(2\rho LS)^2}$). (c) For heavy damping ($b>\sqrt{\omega_0^2(2\rho LS)^2}$), the fluid attempts to return to the equilibrium position slowly.}
    \label{fig:damped}
\end{figure}
Figure~\ref{fig:damped} shows the fluid behavior for different kinds of damping~\cite{the1}. When the damping is light, it holds that $b<\sqrt{\omega_0^2(2\rho LS)^2}$, the fluid oscillates while its amplitude decays exponentially. This system is said to be underdamped, as in the curve (blue).  If the damping constant is $b=\sqrt{\omega_0^2(2\rho LS)^2}$, the system is said to be critically damped, where the fluid does not oscillate and  tends to approach the equilibrium quickly,  as we observe in the curve (red). Curve (green) represents an overdamped system where $b>\sqrt{\omega_0^2(2\rho LS)^2}$ and the  fluid will approach equilibrium over a longer period of time.

\section{Experimental Setup}\label{sececxp}
\begin{figure}[htb]
    \centering
    \includegraphics[width=0.8\linewidth]{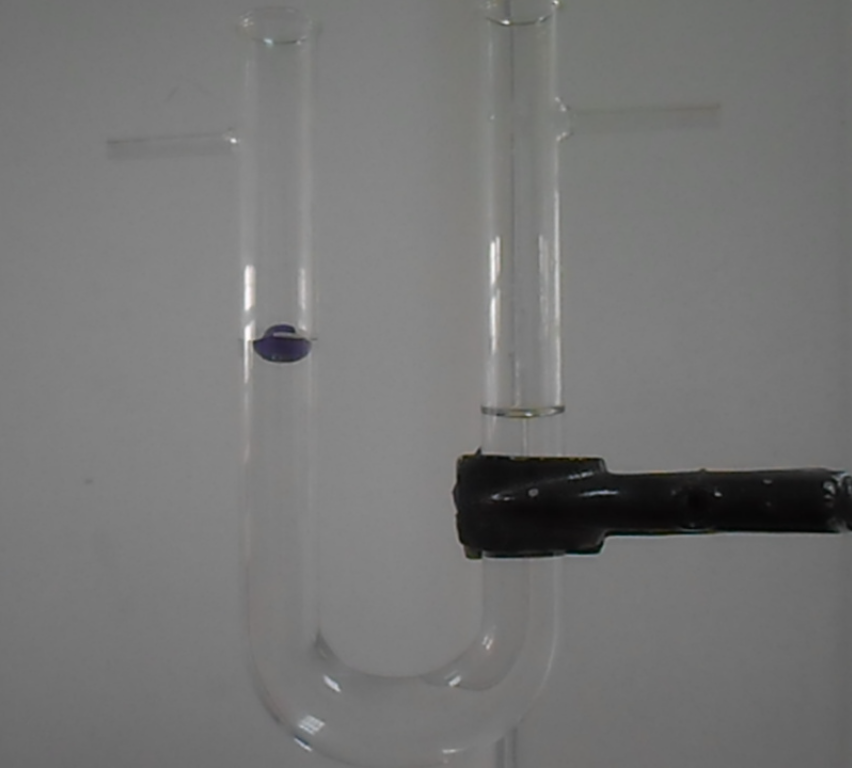}
    \caption{Experimental Setup. }
    \label{fig:lab}
\end{figure}

Figure~\ref{fig:lab} depicts  the experimental setup in the classroom  in order to describe and breakdown any similarities or benefits  between  damped oscillations carried out in physical labs and our simulator. The experiment consists of a U-tube with a constant cross-sectional area $S$. 

\begin{figure}[htb]
    \centering
    \includegraphics[scale=0.55]{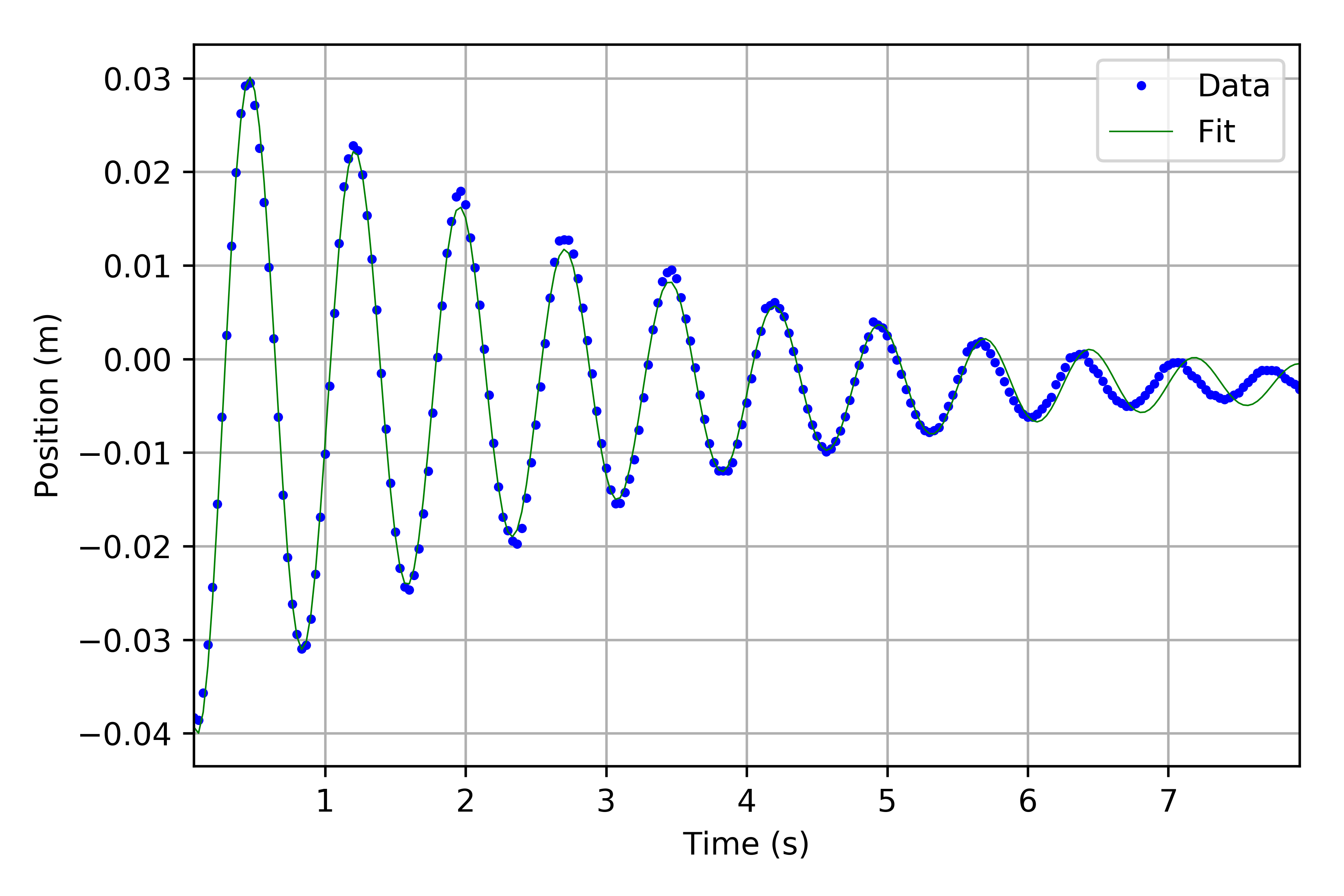}
    \caption{This figure shows the position versus time for small oscillations of a liquid column in the U-tube.}
    \label{fig:pos}
\end{figure}

\begin{figure}[htb]
    \centering
    \includegraphics[scale=0.55]{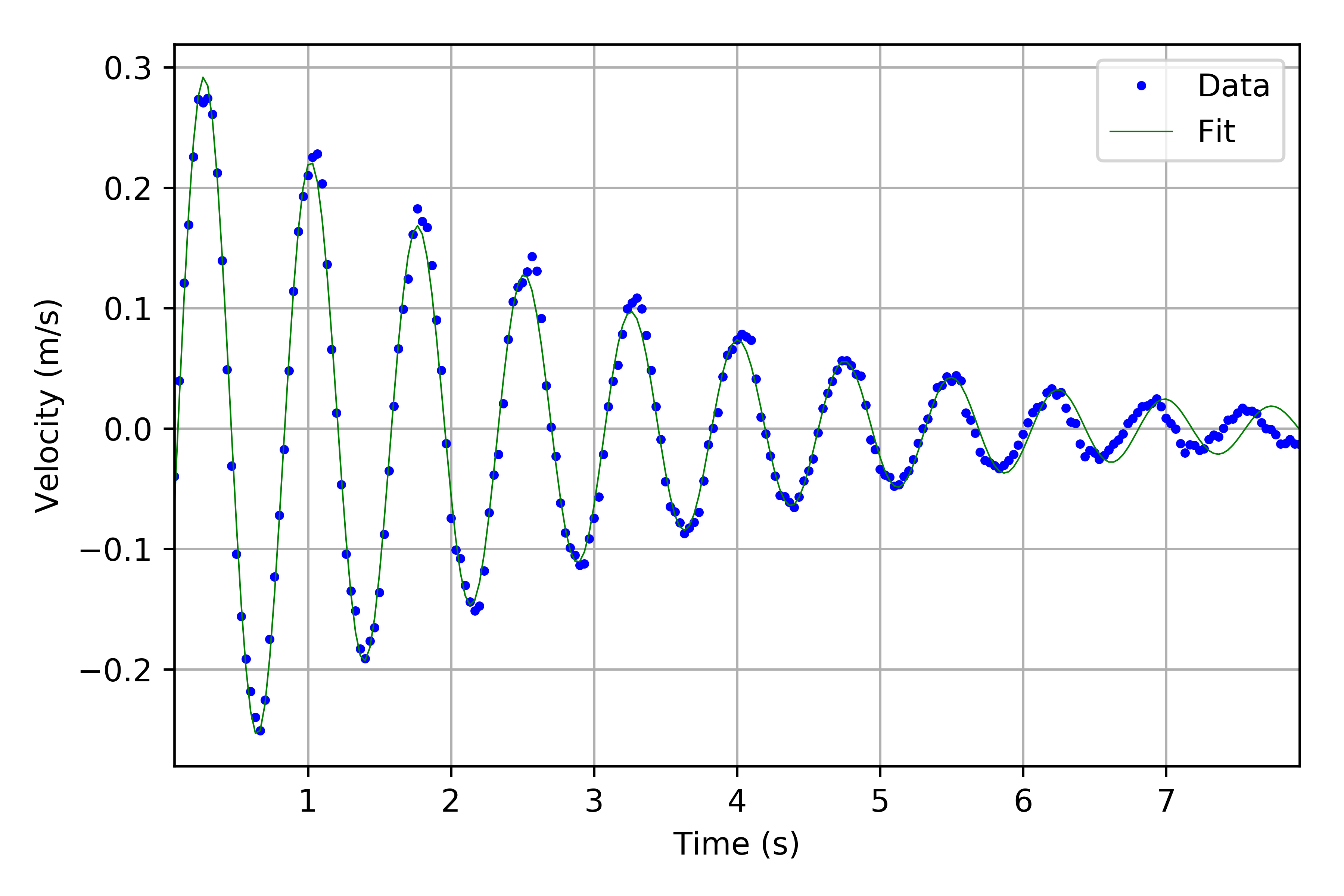}
    \caption{This figure shows the velocity versus time for small oscillations of a liquid column in the U-tube.}
    \label{fig:vel}
\end{figure}

The tube is held at rest relative to the floor  with its
open ends pointing upward. Atmospheric pressure  and  gravitational acceleration  are supposed to be uniform during the experiment. The tube  is partially filled with water~\footnote{Assuming a laminar, incompressible, inviscid, and irrotational liquid.},  and its two  surfaces  mark the same height, which is labeled as the equilibrium state, $y=0$.  After gently blowing one of the tube's arms, the liquid is now set into oscillations back and forth. Tracker video modelling~\cite{tracker} is   a free video analysis and modeling tool built on the Open Source Physics (OSP) Java framework and intended for use in physics education. This tool was used to collect snapshots and csv tables over time intervals in order to track the liquid altitude $y$ and velocity $v_y$ on each side of the tube. In this experiment, we maintained the same parameter configuration and recorded eight  videos (ensembles) suitable for analysis and modelling\href{https://github.com/teaching-physics/liquid-oscillations-Utube}{\faGithub}. The experiment was carried out several times. One of the results is displayed in Figure~\ref{fig:pos} for position, and Figure~\ref{fig:vel} for velocity. 

Points in these plots stand for data obtained from Tracker software, while the continuous lines come from the best-fit parameter found through the model proposed in  Equation~\ref{fit}. From these figures, we can also observe that the experiment configuration describes an underdamped system, while the damping coefficient seems to be very small with respect to the  natural frequency of the oscillations because of the  appreciable number of oscillations. Moreover, when the system is damped, it loses energy during each oscillation, and  we observe the position and velocity decreasing in amplitude as time continues. Nevertheless, the experimental points in Figures~\ref{fig:pos}-~\ref{fig:vel} manifest  a slight deviation  with respect to the theoretical curve after six seconds. We attribute this difference to the assumed linear velocity dependency written in Equation~\ref{eq:Newton1}. The breaking approach  can be based upon  capillary and non-linear viscosity effects that are hard to model in the lab~\cite{the1}.

\begin{figure}[htb]
    \centering
    \includegraphics[scale=0.55]{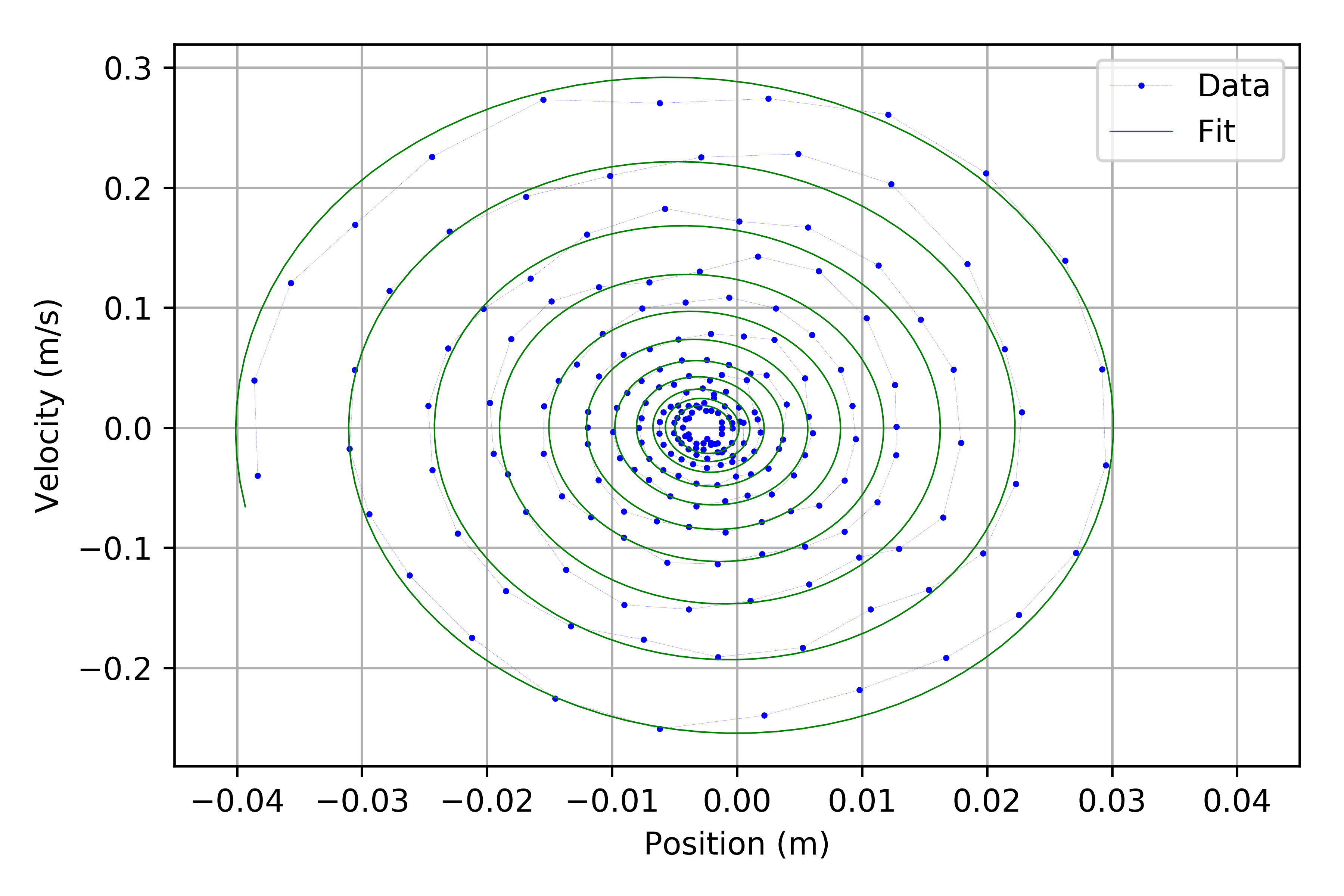}
    \caption{This figure shows the velocity versus position  for small oscillations of a liquid column in the U-tube.}
    \label{fig:phase}
\end{figure}
This behavior is shown in Figure~\ref{fig:phase} where the phase diagram spirals inward due to energy loss. In classical mechanics, this is described as an attractor, causing the system  to be  trapped in a potential well from which it cannot escape. Although it is  clear that this experimental setup can be easily adapted in  physics labs, establishing  alternative configurations  can become costly and even unfeasible. Modifications in the tube specifications, such as  cross-sectional area or length; the use of different fluids; and  changes in gravity are  examples of configuration systems  that can be hard (where characteristics of economics subject can certainly be a priority), or even impossible to analyze in the classroom. Here, simulations will play an important role in opening the window to inaccessible setups. 
\section{Simulation}
\begin{figure}[htb]
    \centering
    \includegraphics[width=\linewidth]{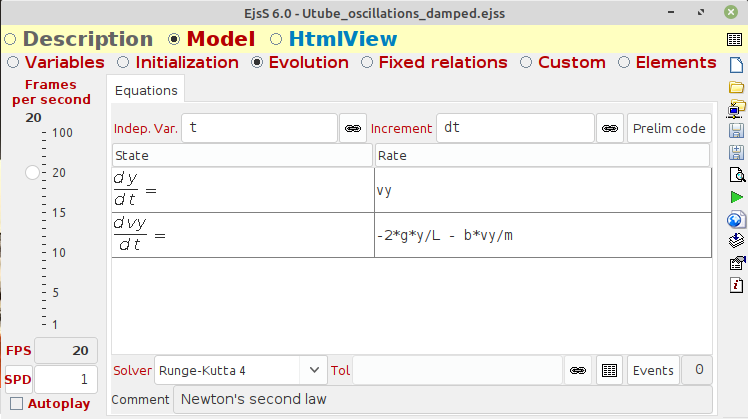}
    \caption{Graphical User Interface (GUI) of the simulation.}
    \label{userinter}
\end{figure}

In this section,  we highlight  our main results  in two steps, the design of an interactive simulation via EJS,  and its use. Easy Java/JavaScript Simulations (EJS) is a free software  created  to support teachers and educational researchers to  implement and deploy computer
simulations in the classroom.  EJS facilitates the numerical approach,
visualization, and deployment of physical simulations so that teachers can focus on the pedagogical aspects rather than programming~\cite{1910}. EJS follows a Model-View-Controller software interface with the purpose of defining the model variables and their equations of motion. Furthermore, the graphical interface allows users to visualize the behavior of the system and interact with it by varying and executing the variables. As mentioned previously, 
the simulation has two aims: to compare with the experimental results described in Section~\ref{sececxp}, and to provide scenarios in which students can work under conditions that are extremely difficult, costly, or time-consuming to create in  physics laboratories. 
EJS contains a variety of interactive tools  that allow one to create pedagogical and interactive Java simulations without the need to be a Java programmer~\cite{ejs1}. The EJS user interface shown in Figure~\ref{userinter} is where the user builds the main aspects related to the simulation.

It consists of three work panels: \textit{Description}, where the user describes detailed information about the simulation; \textit{Model}, where the model is defined. The latter contains various sub-panels, such as \textit{Variables}, which are used to declare and initialize the model's variables; \textit{Initialization}, which  allows additional initialization algorithms to be entered; and \textit{Evolution}, which describes the method  used for solving the  differential equation  of the physical phenomena~\cite{ejs1,ejs2}. 
\begin{figure}[htb]
    \centering
    \includegraphics[width=\linewidth]{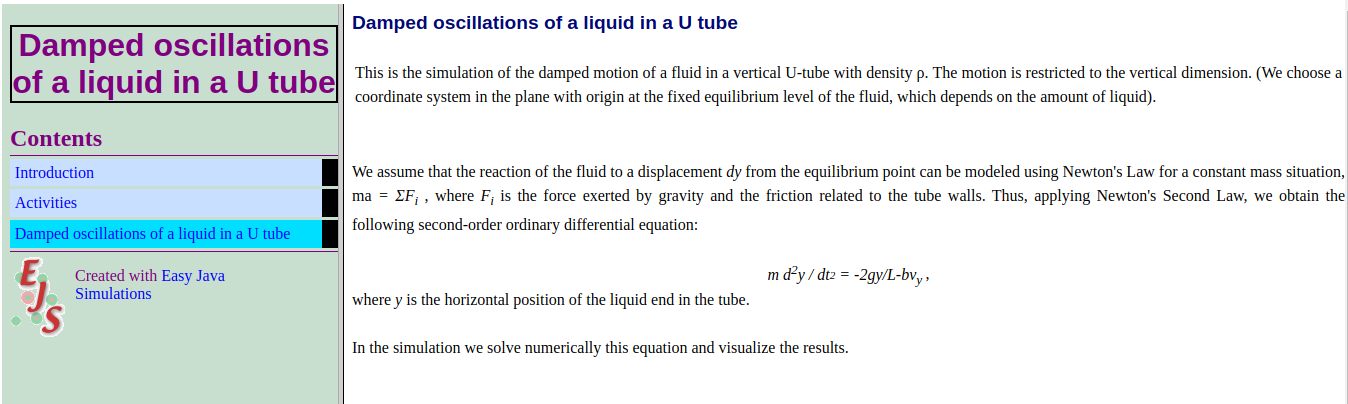}
    \caption{Description of the simulation using the user interface in EJS.}
    \label{desc}
\end{figure}
Finally, the third work panel is called \textit{HtmlView},  which creates the  graphical interface that includes visualization, user interaction, and program control. Figure~\ref{desc} depicts the "Description" work panel developed for the simulation, which includes a brief documentation containing not only the purpose of the simulation but also  a short theoretical framework with the differential equation involved. Regarding the \textit{Model} work panel, the defined variables are related to  the liquid density $\rho$, its damped parameter $b$,  the cross-section area of the tube $S$, and  the liquid column length in both arms, $h_r$ and $h_l$. In addition, initial position $y$ and initial velocity $v_y$ are required as initial conditions for finding solutions to the second-order differential equation shown in Equation~\ref{eq:Newton1}. This equation is solved via the classical fourth-order Runge-Kutta method with a fixed step~\cite{ejs1}. The above description is configured in the user interface, as we see in Figure~\ref{userinter}.
\begin{figure}[htb]
    \centering
    \includegraphics[width=\linewidth]{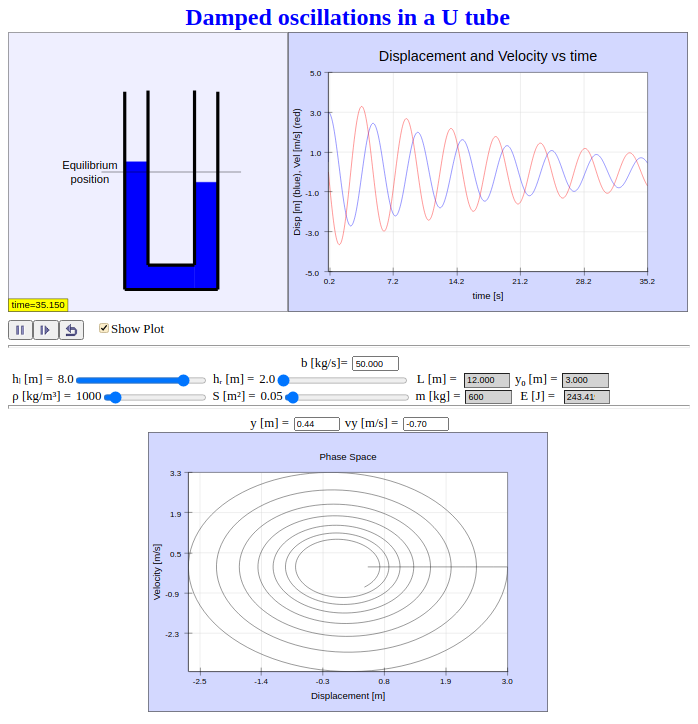}
    \caption{View of the EJS simulation. The upper-left plot displays the interactive U-tube simulator. The time evolution of position and velocity is shown in the upper-right plot. The center plot depicts the interactive buttons to vary the parameters of the experiment. The bottom plot shows the phase diagram.  }
    \label{sim}
\end{figure}
Finally, the \textit{View} work panel is shown in Figure~\ref{sim}.  In order to provide the user with a detailed view of the virtual experiment, three panels are displayed: the time evolution for the parameters, a control panel, and the phase diagram. Both the simulations and the experimental results are available in the public repository \href{https://github.com/teaching-physics/liquid-oscillations-Utube}{\faGithub}. Hence, a more advanced student or instructor can use this material to go one step beyond and modify the simulations to approach more real systems\textbf{\includegraphics[height=10pt,width=10pt]{images.jpeg}}\href{https://teaching-physics.github.io/liquid-oscillations-Utube/}{EJS}.
In order to encourage students or tutors to create their own simulations, they shall first install the EJS which is available for free download from the Open Source Physics website (\href{https://www.compadre.org/osp/items/detail.cfm?ID=12365}{physics}). After launching EJS, they can follow the next steps: 
\begin{enumerate}
    \item To create a new simulation, click on ``New'' in the EJS File menu. This will open a new window where you can set up the basic parameters of your simulation, such as the title, author, and description.
    \item Design the simulation: Once you have set up the basic parameters of your simulation, you can start designing the simulation itself. EJS provides a range of tools and components that you can use to create your simulation, such as buttons, sliders, text fields, and graphs. You can add these components to your simulation by dragging and dropping them onto the simulation window.
    \item Add code: To create a functional simulation, the students should add code to their simulation. Since EJS uses the Java programming language, students will need some basic knowledge of programming. The code is added in a separate tab named "Script" in the EJS main window. Here, they specify the dynamical variables such as positions, velocity,  acceleration, and the equations of motion that evolve the system. EJS include plenty of numerical methods, therefore the student does not need to re-implement them, only would need to provide the initial conditions to solve completely the system of equations specified. 
    \item Test and run the simulation: Once the equations of motions and graphical aspects of the simulation are designed and coded, they can test and run it within EJS. It is as simple as one click on the ``Run'' button in the EJS main window. This will launch your simulation in a web browser.
    \item Optionally, the students can share their simulations with others: EJS allows them to export of the simulation as a Java applet or as a standalone application that can be run on any computer.
\end{enumerate}
Note that this is just a general overview of the process of creating simulations using EJS. Depending on the complexity of the simulation, the student may need to use additional features and tools within EJS. Additionally, it may be helpful to consult EJS documentation or tutorials to learn more about specific features and techniques.

\section{Discussion}
\begin{figure}[htb]
    \centering
    \includegraphics[width=0.5\linewidth]{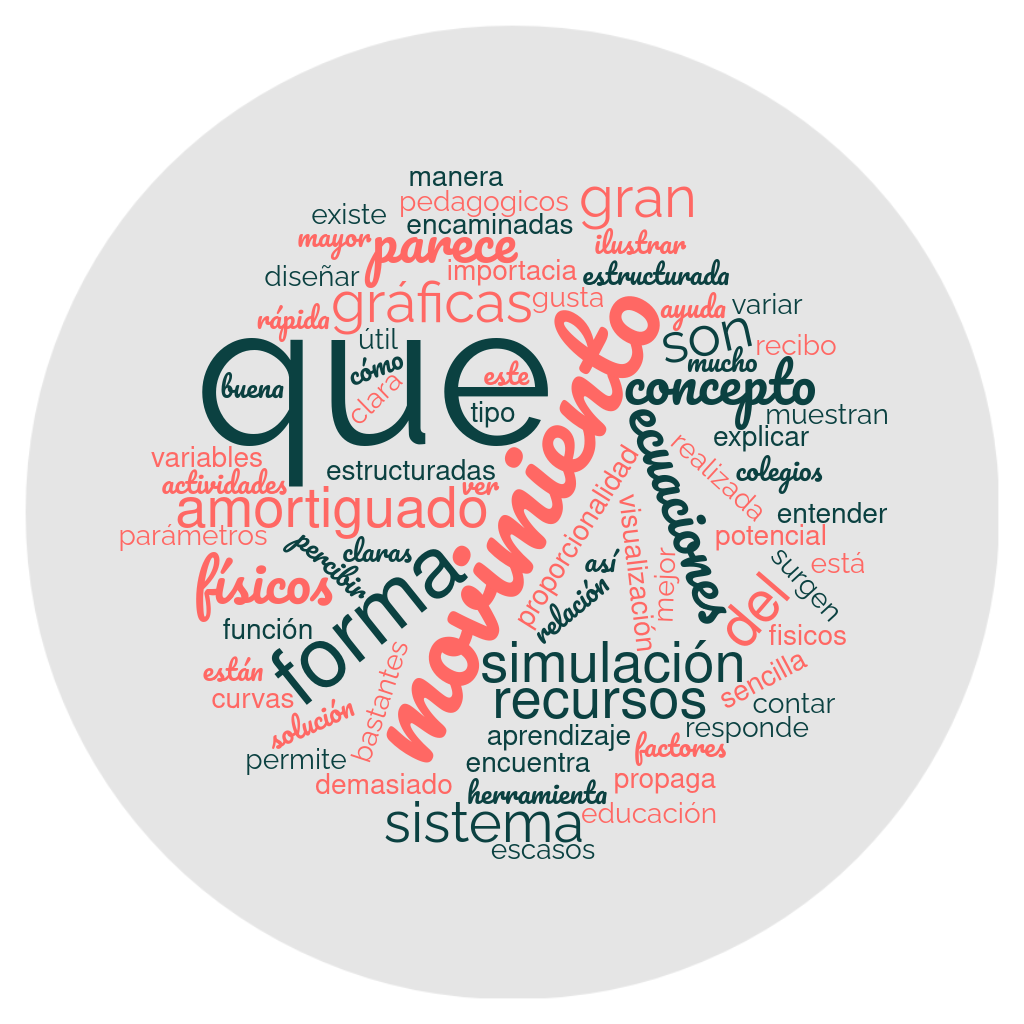}
    
    \caption{Keyboard cloud from student's comments (in  Spanish). Some frequent words are  motion, learning, good, awesome, explainable, easy, and understandable,  among others; with a sentimental analysis around $90\%$ being positive. }\label{fig:comments}
\end{figure}

Simulations and virtual labs offer either an alternative or an  educational supplement to physics laboratory experiments. The simulation presented in this paper might bring benefits when  institutes face practical barriers related to cost, a lack of adequate lab facilities, and distance learning education difficulties. Therefore, with this interactivity feature, students can use the potential of this simulation to design their own lab, by adjusting various parameters and collecting  data to test their hypothesis. However, assessing the acquired experimental skills to gain knowledge on scientific subjects remains challenging to determine, despite being one of the most important laboratory goals~\cite{paxi}.
As previously stated, EJS provides a  suitable frame not only for building physics simulations but also for its  ease of  access. Nevertheless, complicated simulations will also demand a high level  of programming time from educators. Another drawback  of the simulation is the inability to record the data into spreadsheets useful to analyse or visualise their findings.  Still, descriptive science simulator approaches combined with high-quality data will provide students with a  rewarding learning experience~\cite{lye}. Future work should replicate the results of this study with other physics topics, with the goal of replicating EJS simulations  in a broader range of educational institutes and lab modules.\\
Finally, students and teachers from the Vibrations and Waves courses are invited to use the developed simulation in their classes. We hope that this didactic resource will allow students to identify damped oscillations using the fluid in the U-tube, complementing the well-known mass-spring and frictional pendulum systems. Additionally, we expect students to gain a better understanding of the physical laws associated with damped systems by manipulating the simulation's parameters.
In particular, this simulation was used in a Vibrations and Waves course for third-year college physics majors. Students and teachers were asked to complete a survey to collect feedback on the simulation. The following are some of the students' comments (see also the Keyboard cloud in Figure~\ref{fig:comments}):
\begin{itemize}
    \item “The simulation is very well done. It can be seen how the damped oscillations depend on the liquid density”
    \item “It seems to me a very useful tool to illustrate and explain the concept of damped motion. It is well structured and easy to use, which is of great help to better understand the concept and the relationship between variables. The graphs they display are also very well structured and quite clear”.
\end{itemize}

Regarding teachers, some of their opinions about the simulation are as follows:

\begin{itemize}
    \item “I really like the visualization of the graphs. It responds very quickly to the equations of motion of the system. I believe that the opportunities for variation of the system parameters are just right”. 
    \item “It has great potential to design activities aimed at education. The greatest lesson I received from this simulation is perceiving the importance of having this type of educational resources since in many schools physical resources are very scarce”.
\end{itemize}

The use of this type of computational practice as a supplement in physics classes may enhance students' conceptual comprehension of the physical situations being studied. We believe that the ability to observe how the variation of some parameters affects the behavior of a system and, in turn, the ability to recognize the equations as physical relationships between quantities are essential for the understanding of physical phenomena.

\section{Conclusions}
In this paper we introduced the implementation details of the simulation based on Easy Java/JavaScript ,Simulations (EJS) to study the problem of damped oscillations in U-tubes\textbf{\includegraphics[height=10pt,width=10pt]{images.jpeg}}\href{https://teaching-physics.github.io/liquid-oscillations-Utube/}{EJS}. We also highlighted various advantages associated with EJS's current capabilities in terms of simulation design and accessibility for teaching science, making it an increasingly popular online learning platform. We expect that teachers will find this helpful simulation as a supplement to physics laboratory experiments.


\section*{References}
\begin{thebibliography}{9}
\bibitem{Monther}A. Monther, \textit{Evaluating virtual laboratory platforms for supporting on-line information security courses}, doi.org/10.48550/arxiv.2208.12612 (2022) 
\bibitem{Rahman}Rahman, Fozlur and Mim, Marium Sana and Baishakhi, Feekra Baset and Hasan, Mahmudul and Morol, Md. Kishor, \textit{A Systematic Review on Interactive Virtual Reality Laboratory}, Proceedings of the 2nd International Conference on Computing Advancements,\textbf{10}, 491 (2022) 
\bibitem{Elmqaddem} N. Elmqaddem, \textit{Virtual reality in education. Myth or reality?},International Journal of Emerging Technologies in Learning,\textbf{14}, 3 (2019) 
\bibitem{Hut} P.Hut, \textit{Virtual laboratories and virtual worlds}, Dynamical Evolution of Dense Stellar Systems
Proceedings IAU Symposium,\textbf{3},S246 (2007) 
\bibitem{Clemente} F.G. Clemente, F. Esquembre, \textit{Deployment of physics simulation apps using Easy JavaScript Simulations}, 2017 IEEE Global Engineering Education Conference (EDUCON), \textbf{1},1093 (2017) 
\bibitem{Wee} L. K.Wee, \textit{Physics Educators as Designers of Simulation using Easy Java Simulation (Ejs)}, American Association of Physics Teachers National Meeting Conference, 91 (2010) 
\bibitem{Cukurova} M. Cukurova,  J. Bennett, A. I. Z. Abrahams, \textit{Students’ knowledge acquisition and ability to apply knowledge into different science contexts in two different independent learning
settings}, Informa UK limited, trading as Taylor and Francis group, (2017)
\bibitem{Laru} P. Näykki, J. Laru , E. Vuopala, P. Siklander  and S. Järvelä  Affective \textit{Learning in Digital Education—Case Studies of Social Networking Systems, Games for Learning, and Digital Fabrication}, Front. Educ. \textbf{4}, 128 (2019) 

\bibitem{OSP} \textit{Open Source Physics, OSP Collection}, http://www.opensourcephysics.org/.
\bibitem{phet} \textit{University of Colorado, PhET Interactive Simulations}, https://phet.colorado.edu/.
\bibitem{fisica} \textit{Fisica por ordenador}, http://www.sc.ehu.es/sbweb/fisica/
\bibitem{ejs1} \textit{University of Murcia, Easy JavaScript Simulation}, http://www.um.es/fem/EjsWiki/
\bibitem{tracker} \textit{Tracker video modelling}, https://physlets.org/tracker/
\bibitem{ejs2} F. J. Garcia Clemente, F. Esquembre, \textit{Ejss: A javascript library and
authoring tool which makes computational-physics education simpler}, XXVI IUPAP Conference on Computational Physics (CCP), Boston, USA, (2014)
\bibitem{Jorge} J.E. Garc\'ia-Farieta, J. Salamanca and D. Rodríguez, \textit{Radiación-Materia: Geant4 Hands On! Un recurso educativo orientado a profesores-investigadores}, (Universidad Distrital Francisco José de Caldas, Bogota, 2020), 1ª ed.
\bibitem{the1} A. Ogawa et.al, \textit{Damped oscillation of liquid column in vertical U-tube for newtonian and
non-newtonian liquids}, Journal of Thermal Science \textbf{16}, 4 (2007)
\bibitem{the2} H. M. Aguilar et.al, \textit{Liquid oscillations in a U-tube}, Phys. Educ.  \textbf{53}, 015005 (2018)

\bibitem{1910} F. C. Esquembre, F.L. Garc\'ia  et.al, \textit{Easy Java/JavaScript Simulations as a tool for Learning Analytics}, arXiv1910.09156 (2019)

\bibitem{paxi} E. Paxinou   et.al, \textit{Assessing the impact of virtualizing physical labs}, Conference: EDEN 2018 ANNUAL Conference, Exploring the Micro, Meso and Macro Navigating between dimensions in the digital learning landscape, (2018)
\bibitem{lye} S. Y. Lye et.al, \textit{Design, customization and implementation of energy simulation with 5E model in elementary classroom}, Educational Technology and Society  \textbf{17}, 3 (2014)
\end{thebibliography}
\end{document}